\documentclass[aps,pra,reprint,showpacs,groupedaddress]{revtex4-1}
\usepackage{graphicx}
\usepackage{amsmath,amssymb}
\usepackage{bm}

\begin{document}


\title{Photoionization loss in
simultaneous magneto-optical trapping of Rb and Sr}


\author{Takatoshi Aoki}
\altaffiliation[
Email address: ]{aoki@phys.c.u-tokyo.ac.jp}
\affiliation{Institute of Physics,
Graduate School of Arts and Sciences,
University of Tokyo, Tokyo 153-8902, Japan}

\author{Yuki Yamanaka}
\affiliation{Institute of Physics,
Graduate School of Arts and Sciences,
University of Tokyo, Tokyo 153-8902, Japan}
\author{Makoto Takeuchi}
\affiliation{Institute of Physics,
Graduate School of Arts and Sciences,
University of Tokyo, Tokyo 153-8902, Japan}
\author{Yasuhiro Sakemi}
\affiliation{Cyclotron and Radioisotope Center, Tohoku University,
Sendai 980-8578, Japan}
\author{Yoshio Torii}
\affiliation{Institute of Physics,
Graduate School of Arts and Sciences,
University of Tokyo, Tokyo 153-8902, Japan}

%
%

\date{\today}

\begin{abstract}

We demonstrate the simultaneous magneto-optical trapping (MOT) of Rb
and Sr and examine the characteristic loss of Rb in the MOT due to
photoionization by the cooling laser for Sr. The photoionization
cross section of Rb in the $5P_{3/2}$ state at 461 nm is determined
to be $1.4(1)\times10^{-17}$ cm$^2$. It is important to consider this
loss rate to realize a sufficiently large number of trapped Rb atoms
to achieve a quantum degenerate mixture of Rb and Sr.

\end{abstract}

\pacs{32.80.Fb, 67.85.-d, 37.10.Gh}


%
\maketitle


\section{Introduction}

During the last decade, quantum degenerate mixtures of different
atomic species have been investigated by many researchers
\cite{LiNa,LiK,LiRb,KNa,KRb,RbCs}.
These systems have included research of
Bose--Fermi mixtures \cite{Sengstock}, the
Fulde--Ferrell--Larkin--Ovchinnikov (FFLO) state near the
BCS--Bose--Einstein--condensate (BEC) crossover
with different masses \cite{FFLO}, self-trapping in optical lattices
\cite{self trapping}, and the heteronuclear Effimov state
\cite{Effimov}. The heteronuclear molecule comprised of two
different atomic species in the rovibrational ground state has an
electric dipole moment, and ultracold RbK molecules in the
rovibrational ground state have been realized using stimulated Raman
adiabatic passage (STIRAP) \cite{JinRbK}. Ultracold polar molecules
enable us to study new quantum phases \cite{crystaline} and quantum
logic gates \cite{DeMille}.

Recently, polar molecules comprised of alkali and alkali-earth (or
rare-earth) atoms have come under increasing scrutiny because such
polar molecules have an electron spin in the rovibrational ground
state in addition to an electric dipole moment, which offers the
chance to study new quantum phases of the lattice spin model
\cite{LatticeSpin}, precise measurement \cite{Kajita}, and
fundamental physics \cite{aokiFPUA}. One such example is YbRb, whose
molecules in the excited states are generated by two-photon
photoassociation in a mixture of ultracold Rb and Yb \cite{YbRb2PA}.
Recently, quantum degenerate mixtures of LiYb have been reported
\cite{LiYb}, but the LiYb molecule has a relatively small electric
dipole moment (0.02--0.15 Debye) \cite{LiYbtheory}. In contrast, the
RbSr molecule is expected to have a relatively large electric dipole
moment (1.4 Debye) \cite{ASr}. Furthermore, the theoretically
predicted heteronuclear Feshbach resonances of RbSr \cite{RbSr}
indicate that it is possible to associate ultracold Rb and Sr atoms
to produce the RbSr molecule. Because techniques for laser cooling of Rb
and Sr are well established, quantum degeneracy of Sr has been
achieved by several groups \cite{84SrBEC}. However, simultaneous
laser cooling of Rb and Sr has not yet been reported.

\begin{figure}[!b]
\begin{center}
\includegraphics[clip, width=1\columnwidth]{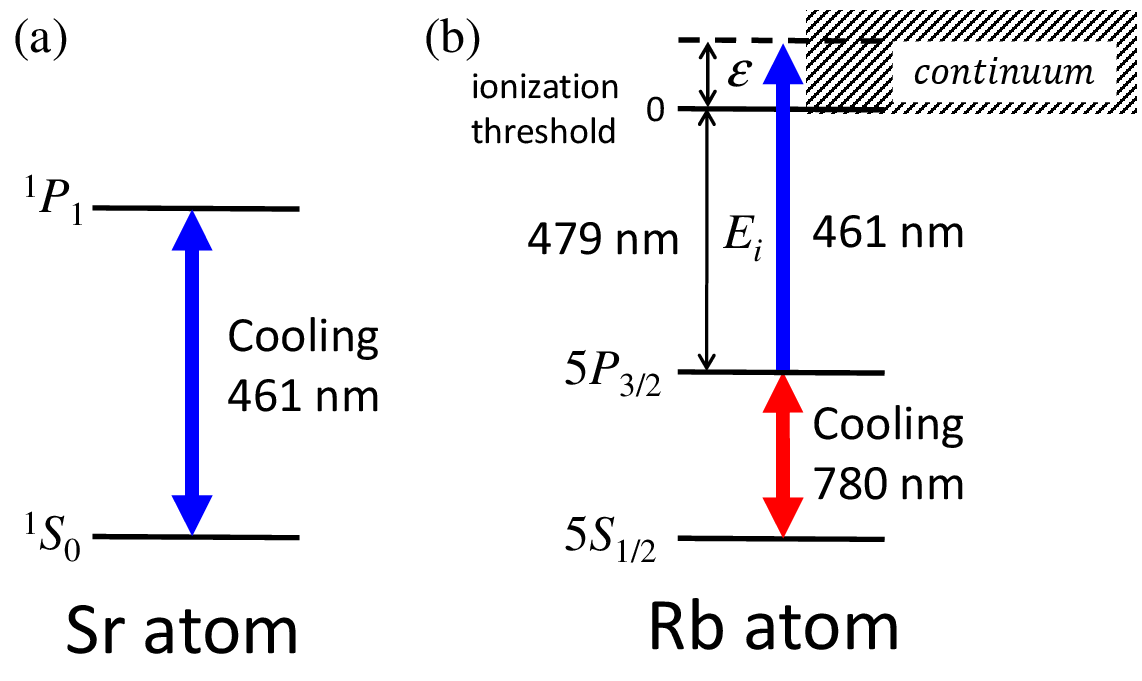}
\end{center}
\caption{(Color online)
Energy diagrams and relevant transitions of (a) Sr and (b)
Rb. The ionization energy $E_{i}$ from the $5P_{3/2}$ state of Rb
corresponds to a wavelength of 479 nm. Excess energy $\epsilon$ is
converted into the kinetic energy of the ionized Rb atom and the
electron. } \label{fig:photoionization}
\end{figure}

In this work we demonstrate simultaneous magneto-optical trapping
(MOT) of Rb and Sr. We observe the characteristic loss of Rb in
the MOT, which can be understood as photoionization due to the
461-nm cooling beam for the Sr atoms. The loss rate was measured by
changing the intensity of the 461-nm laser beam, and the
photoionization cross section of Rb in the $5P_{3/2}$ state was
determined. We found that simultaneous MOT of Rb and Sr is possible
with a cooling beam for Sr at the saturation intensity.

\section{Photoionization in simultaneous MOT}

Photoionization is the ionization of a neutral atom due to an
atom--light interaction. Figure~\ref{fig:photoionization} shows the
energy diagrams and laser transitions for simultaneous laser cooling
of Rb and Sr. The wavelengths of the cooling transitions are 461 and
780 nm for Sr and Rb, respectively. The 780-nm laser does not ionize
Sr in the excited state $^{1}P_{1}$ because photoionization from
this state requires a photon energy of 415 nm. However, the
wavelength from the Rb $5P_{3/2}$ state to the ionization threshold
is 479 nm, and so the 461-nm laser can ionize Rb atoms in the
$5P_{3/2}$ state. This photoionization would result in trap loss in
simultaneous MOT of Rb and Sr. For previously reported
alkali--metal--alkali--metal mixtures, i.e., Li--Na \cite{LiNa}, Li--K \cite{LiK},
Li--Rb \cite{LiRb}, K--Na \cite{KNa}, K--Rb \cite{KRb}, Rb--Cs
\cite{RbCs}, and Yb--Li \cite{LiYb}, photoionization due to the
cooling laser did not occur because the photon energies of the
cooling beams were smaller than the photoionization energies.

The loss rate $R$ due to
photoionization is given by \cite{Florian,photo},

\begin{equation}
R = \Phi f \sigma,
\label{loss}
\end{equation}

\noindent where $\Phi = I/h\nu$ is the photon flux, $I$ is the
intensity of the photoionization laser beam, $h$ is the Planck
constant, $\nu$ is the optical frequency of the photoionization
laser beam, $f$ is the population fraction of the excited state,
i.e., the $5P_{3/2}$ state of Rb, and $\sigma$ is the
photoionization cross section. Taking into account the
photoionization loss rate, the rate equation for the number of Rb
atoms in the MOT is written as \cite{photo2}

\begin{eqnarray}
\frac{dN_{Rb}(t)}{dt} &=& \Gamma - \frac{1}{\tau} N_{Rb}(t) - B - R N_{Rb}(t),
\label{N1}
\end{eqnarray}

\noindent where the $N_{Rb}(t)$ is the number of trapped Rb atoms as
a function of time, $\Gamma$ is the loading rate of the Rb MOT, and
$\tau$ is the decay time constant due to background gas collisions.
The loss coefficient $B = \beta_{RbRb} \int n_{Rb}^{2}(r) d^3r +
\beta_{RbSr} \int n_{Rb}(r) n_{Sr}(r) d^3r$ is due to light-assisted
collisions, where $\beta_{RbRb}$ and $\beta_{RbSr}$ are the
light-assisted collision coefficients between the Rb atoms and
between the Rb and Sr atoms, respectively, and $n_{Rb}(r)$ and
$n_{Sr}(r)$ are the respective spatial densities of Rb and Sr.
Because light-assisted collisions between the Rb and Sr atoms were
not observed, as described in section IV, and the density of the
trapped Rb can be assumed to be constant in a usual MOT experiment \cite{density_constant},
$B$ can be approximated as $B \simeq \beta_{RbRb} n_{Rb} N_{Rb}(t)$,
where $n_{Rb}$ is a constant density. 
Then, Eq. (\ref{N1}) is written as

\begin{eqnarray}
\frac{dN_{Rb}(t)}{dt} &=& \Gamma - \frac{1}{\tau'} N_{Rb}(t) - R N_{Rb}(t),
\label{N2}
\end{eqnarray}

\noindent where $1/\tau' \equiv 1/\tau + \beta_{RbRb} n_{Rb}$. 
The solution to
Eq. (\ref{N2}), when $R$ $(\neq0)$ is introduced at time $t$ = 0, is
written as

\begin{equation}
N(t) = N_{\infty} + (N_{0}-N_{\infty}) \exp(-R't),
\label{N3}
\end{equation}

\noindent where $R' \equiv R + 1/\tau'$,  
$N_{0}$ is the steady state atom number for the rate
equation with $R$ = 0, and $N_{\infty}$ is that for an arbitrary
$R$. Using this solution, measurement of the decay of the number of
Rb atoms in the MOT allows us to determine the photoionization rate
$R$ and the photoionization cross section $\sigma$ of Rb in the
$5P_{3/2}$ state at 461 nm.

\begin{figure}[t]
\includegraphics[clip, width=1\columnwidth]{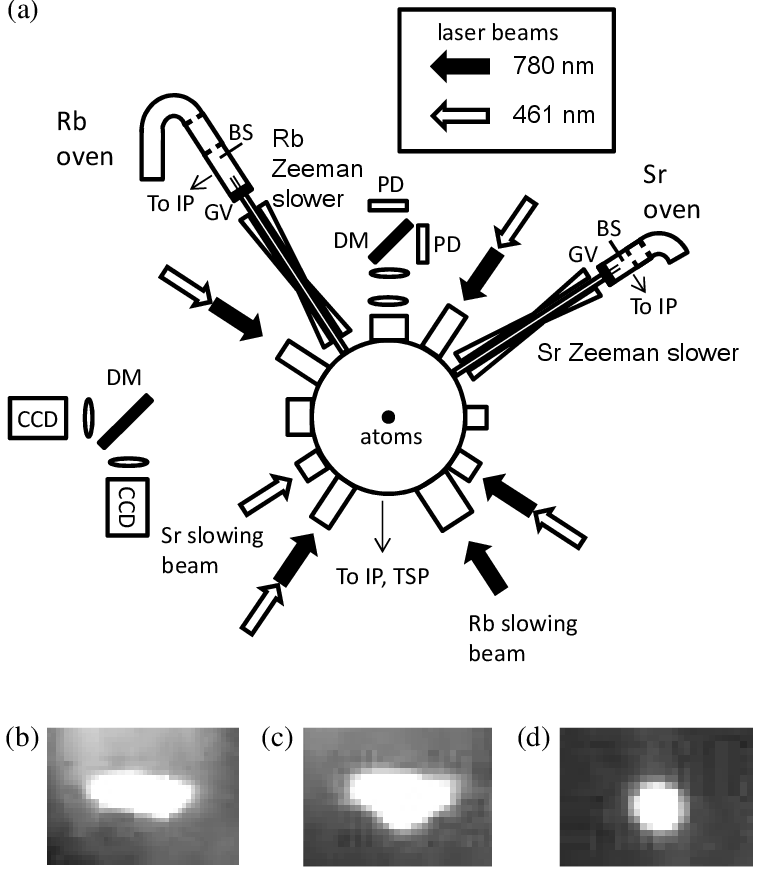}
\caption{(a) Schematic diagram of the experimental apparatus. IP:
ion pump; TSP: Ti-sublimation pump (these vacuum pumps are not
shown); BS: beam shutter; GV: gate valve; DM: dichroic mirror; and
PD: photo detector. The 780 and 461 nm laser beams are shown by the
black and white arrows, respectively. (Trapping beams propagating
perpendicular to this diagram are not shown.) Fluorescence images of
the trapped (b) Sr, (c) Sr and Rb, and (d) Rb atoms. In (c), the
trapped clouds of Rb and Sr are spatially overlapped. The field of
view is 7.5 mm $\times$ 5.6 mm. }
 \label{fig:MOT}
\end{figure}

\section{Simultaneous MOT}

Our laser system and frequency locking scheme for cooling the Sr
atoms are similar to those described in Ref. \cite{Aoki}. Briefly, a
laser beam with a wavelength of 922 nm from an extended cavity diode
laser (ECDL) was amplified by a tapered amplifier (TA) up to 900 mW.
This laser beam enters a frequency doubling cavity, which generates
a 461-nm laser beam with a power of 175 mW. The frequency was
stabilized by using frequency modulation spectroscopy with a Sr
hollow-cathode lamp. The repumping beam, also generated by frequency
doubling, has a wavelength of 497 nm and is resonant with the
transition between 5s5p$^{3}P_{2}$ and 5s5d$^{3}D_{2}$ states.

For cooling the Rb, we used a commercially available extended cavity
tapered laser (ECTL), and an ECDL was used for the repumping. The
frequencies of these laser beams were stabilized by the polarization
spectroscopy of a Rb cell \cite{Torii}.

Figure 2 (a) shows the vacuum system, which was comprised of ovens,
Zeeman slowers, and a main chamber with a design similar to that in
Ref. \cite{MITRbBEC machine}. The vapor pressures of Rb and Sr are
quite different: a pressure of $1 \times 10^{-4}$ Torr inside the
oven is achieved at about 100 $^\circ$C for Rb but at 400 $^\circ$C
for Sr. We therefore prepared separate ovens and Zeeman slowers for
each atomic species. The Rb (Sr) atomic beam was extracted from the
Rb (Sr) oven and passed through the oven chamber, whose background
pressure was $10^{-8}$ Torr, whereas the background pressure of the
main chamber was kept below $2\times10^{-11}$ Torr by using a 75-L/s
ion pump and a Ti-sublimation pump.

The Rb atomic beam was first extracted from the Rb oven and
decelerated by the Rb Zeeman slower; the Sr atomic beam is
decelerated by the Sr Zeeman slower. The Zeeman slowers have
zero-crossing designs, and the magnetic field at the exit of the
slower is 100 G (600 G) for Rb (Sr). The trapping beam for the Rb
(Sr) with a wavelength of 780 nm (461 nm) was split into three beams by
waveplates and polarization beam splitters. These beams are
overlapped by dichroic mirrors and pass through the achromatic
quarter waveplates, creating circularly polarized beams. The
repumping beam for Rb is overlapped with the Rb slowing beam,
whereas the repumping beam for Sr is overlapped with one of Sr
trapping beams. The decelerated atomic beams enter the main chamber
and reach the intersection of the three orthogonal trapping beams
for Rb and Sr. The Rb and Sr atoms are then simultaneously trapped.

\begin{figure}[t]
\begin{center}
\includegraphics[clip, width=1\columnwidth]{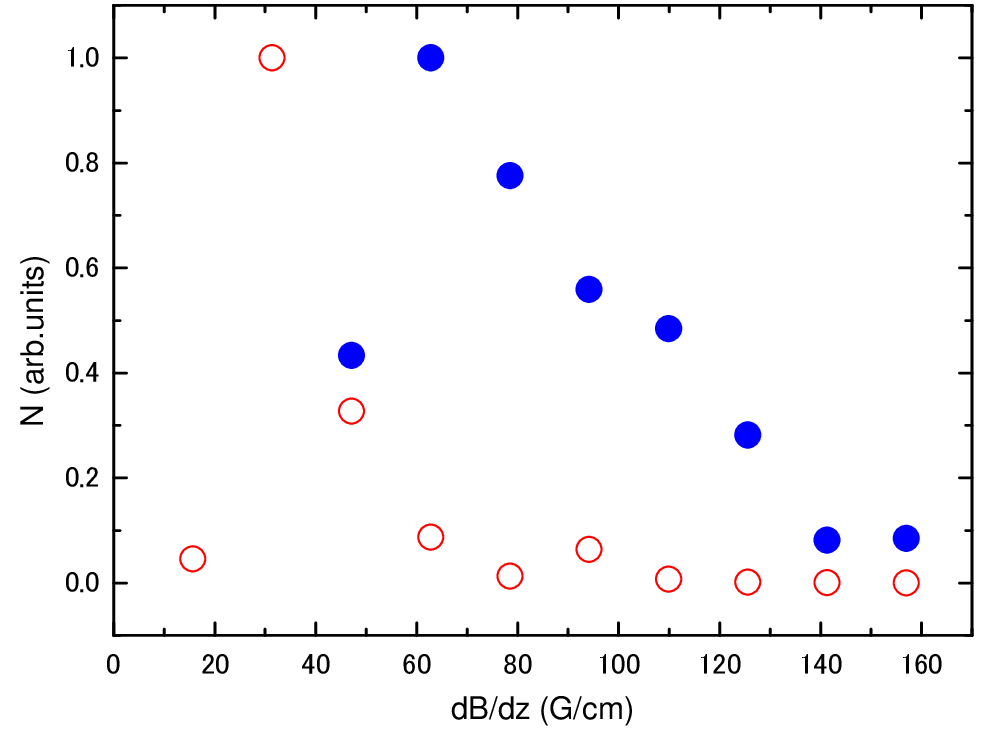}
\end{center}
\caption{(Color online)
Number of trapped Rb and Sr atoms versus the magnetic field
gradient. The solid circles denote Sr, and the open circles denote
Rb. The numbers of trapped atoms are normalized to the maximum
values. } \label{fig:N vs dBdz}
\end{figure}

Figures~\ref{fig:MOT}(b)-2(d) shows fluorescence images of the
trapped clouds in the MOT taken by a charge coupled device (CCD)
camera. The trapped Rb and Sr clouds are spatially overlapped in the
image. The power and diameter of each trapping beam was 15 mW (10
mW) and 34 mm (16 mm), respectively, for Rb (Sr), and the power of
the slowing and repumping beams were 25 mW (2.3 mW) and 20 mW (150
$\mu$W) for Rb (Sr), respectively. A 20-MHz (40-MHz) detuning was
used for Rb (Sr). To obtain an image of both Rb and Sr, we chose a
magnetic field gradient along the axial direction of 90 G/cm to
reduce the size of the trapped Rb cloud to that comparable with the Sr
cloud. The fluorescence from the Rb atoms is stronger than that from
the Sr atoms, and the CCD camera has a higher sensitivity at 780 nm
than at 461 nm. Thus, the fluorescence signal on the CCD camera from
the trapped Rb atoms saturates easily. To reduce the Rb signal, the
fluorescence from both species was passed through a dichroic mirror
that is transparent at 461 nm but reduces the intensity of 780-nm
light. This allowed the fluorescence from both the Rb and Sr to be
simultaneously detected as shown in Fig. 2 (b)-(d).

The dependence of the number of trapped atoms on the magnetic field
gradient is shown in Fig.~\ref{fig:N vs dBdz}. The number of trapped
Rb atoms is maximized ($1 \times 10^{10}$) at a field gradient of 30
G/cm, whereas the number of Sr atoms is maximized ($1 \times
10^{6}$) at 60 G/cm. The difference in the two magnetic field
gradients stems from the difference in the natural line widths (6 MHz for
Rb and 32 MHz for Sr).

\section{Measurement of the photoionization rate}

Understanding the loss is important for achieving a large number of
trapped atoms. Light-assisted collisions have a large effect on
conventional alkali-atom trapping experiments. However, we did not
observe this type of loss in the simultaneous MOT under our
experimental conditions. Instead, when the trapped Rb atoms were
irradiated by the 461-nm laser beams, the number of Rb atoms
decreased due to photoionization.

\begin{figure}[t]
\begin{center}
\includegraphics[clip, width=1\columnwidth]{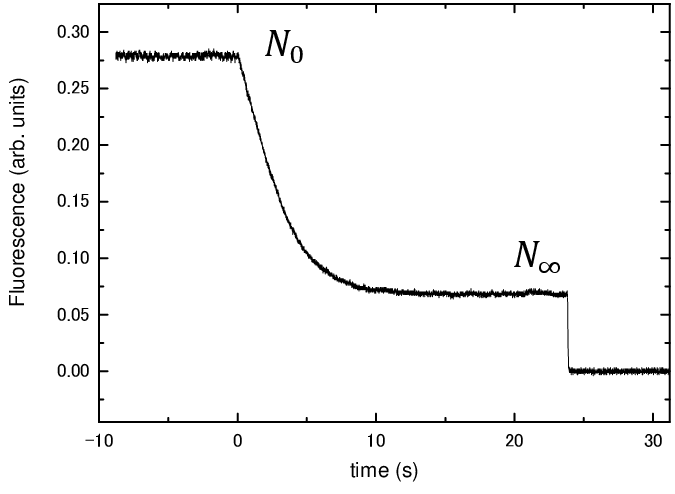}
\end{center}
\caption{Typical decay signal of Rb atoms in the MOT. The 461-nm
laser beam is turned on at $t$ = 0. The intensity of the 461-nm beam
was 307 mW/cm$^2$, which corresponds to an intensity 7.3 times the
saturation intensity for the cooling transition of Sr. }
\label{fig:decay of Rb}
\end{figure}


To measure the photoionization cross section, the trapped Rb cloud
is irradiated by the 461-nm laser beam at various intensities. The
counter propagating beam, which is generated by a mirror, also
interacts with the atoms to double the intensity of the
photoionization beam. We first load a Rb MOT with a trapping beam
diameter of 17 mm, and the fluorescence signal from the trapped Rb
atoms is detected by a photodetector. 
The number of atoms in the MOT exhibits a nearly simple exponential growth 
with a time constant of 5 s, indicating that $1/\tau'\simeq 0.2$. 
After the
the number of trapped atoms saturates, 
the Rb cloud is irradiated by the 461-nm beam.
Figure~\ref{fig:decay of Rb} shows the typical decay due to
photoionization of the number of trapped Rb atoms. The intensity of
the 8-mm photoionization beam was 307 mW/cm$^2$. The number of Rb
atoms decays after the 461-nm beam is turned on, and after the
trapping beams are switched off at 24 s, the signal returns to the
background level of the photodetector.

Figure~\ref{fig:decay of Rb} indicates that trapped atoms remain
after the photoionization-induced decay. The number of trapped atoms
is determined by the balance between the Rb loading rate from the
slowed atomic beam and the photoionization loss rate [Eq. (2)].
If we choose a beam intensity that is lower than
the saturation intensity $I_s$ = 42 mW/cm$^2$, a Rb atom loss of
only 10--20 \% is seen.
A typical trapping experiment requires the
intensity of the MOT beams to be close to that of $I_s$, and thus
this result strongly supports the possibility of simultaneous MOT of
Rb and Sr with a large number of atoms.

\begin{figure}[t]
\includegraphics[clip, width=1\columnwidth]{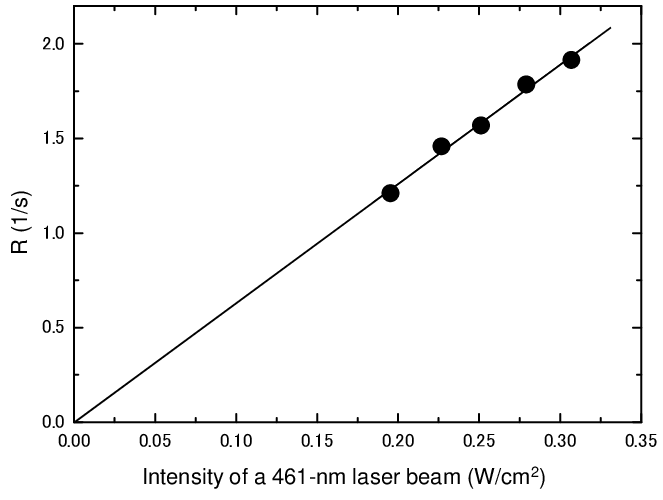}
\caption{Decay constant versus the
intensity of the 461-nm beam. The solid circles are the measured
values, and the solid line is a theoretical fit to the data. }
\label{fig:ionized Rb2}
\end{figure}

The decay constant $R'$ is evaluated by fitting Eq. (4) to the decay
curve.
Then, we obtained $R$ by subtracting $1/\tau'$ from $R'$.
According to Eq. (1), $R$ is
a function of the intensity $I$, and to estimate the photoionization
cross section, we measured $R$ by changing $I$ as shown in
Fig.~\ref{fig:ionized Rb2}. As expected, $R$ is proportional to
$I$. By fitting Eq. (1) to the experimental data and using $f =
0.19$, the photoionization cross section of Rb in the $5P_{3/2}$
state is given as $(1.4 \pm 0.1) \times 10^{-17}$ cm$^2$. The value
of $f$ is calculated from the detuning of the Rb MOT beam ($-$16
MHz) and the intensity $I_{Rb}$ = 12 $I_{Rb,s}$, where $I_{Rb}$ is
the intensity of the Rb MOT beam, and $I_{Rb, s}$ = 1.6 mW/cm$^2$ is
the saturation intensity of the cooling transition for Rb. The
uncertainty in the cross section comes mainly from the uncertainty
in the determination of the population $f$ in the $5P_{3/2}$ state.

The photoionization cross section is determined by the matrix
element of the transition dipole moment between the initial atomic
state and the continuum state above the ionization threshold, and
thus it depends on the excess energy $\epsilon$ in Fig. 1 (b) (see
Ref. \cite{Stewart}). The photoionization cross section of the Rb
atom in the $5P_{3/2}$ state has been investigated by several other
groups for different wavelengths. Aymar {\it et
al.}~\cite{phototheory} calculated the cross section theoretically
and found cross sections of 1.19, 1.22, and 1.30 $\times 10^{-17}$
cm$^2$ at 440 nm, and 1.25, 1.30, and 1.40 $\times 10^{-17}$ cm$^2$
at the 479-nm threshold. Dinneen {\it et al.}~\cite{photo} measured
the cross section to be $1.25(11)$ and $1.36(12) \times 10^{-17}$
cm$^2$ at 407 and 413 nm using MOT of Rb. A value of $1.48(22)
\times 10^{-17}$ cm$^2$ at 476.5 nm was measured by Gabbanini {\it
et al.}~\cite{photo2}, and a cross section of $0.54(12)$ and
$1.34(16) \times 10^{-17}$ cm$^2$ at 296 nm using a pulse laser and
at 421 nm using a continuous-wave laser were found by Ciampini {\it
et al.}~\cite{photoBEC}. Work by Nadeem {\it et
al.}~\cite{photo2011} determined cross sections of 1.25, 1.26, 1.36,
1.5, and 1.88 $\times 10^{-17}$ cm$^2$ at 425, 440, 460, 476.5, and
479 nm using a pulsed laser and a Rb heat pipe with an Ar buffer
gas. The overall uncertainty in the cross section by Nadeem {\it et
al.} is 16 \%, and a cross section at 460 nm of 1.36(21) $\times
10^{-17}$ cm$^2$ was given in Ref. \cite{photo2011}. Our value of
$1.4(1) \times 10^{-17}$ cm$^2$ at 460.862 nm is in good agreement
with this value.

\section{Conclusion}

We have demonstrated the simultaneous MOT of Rb and Sr. Both species
have different optimum values of the magnetic field gradient that
maximize the number of trapped atoms. Loss due to light-assisted
collisions between Rb and Sr was not observed under our experimental
conditions, but the cooling beam for Sr causes
photoionization-induced loss of Rb in the excited $5P_{3/2}$ state.
This had not previously been observed for other alkali and
alkali-earth (two-electron) atom mixtures.

In spite of the photoionization loss, we realized simultaneous MOT
of Rb and Sr under the typical intensity of the 461-nm beam. The
measurement of the photoionization loss rate as a function of the
beam intensity gave the photoionization cross section at 460.862 nm
as $1.4(1) \times 10^{-17}$ cm$^2$. We believe that this value is
useful for optimizing the numbers of trapped Rb and Sr atoms in a
simultaneous trapping experiment.

Currently, we are preparing a laser system for cooling Sr using the
689-nm narrow line transition. The next step is to cool Sr atoms
below 1 $\mu$K and to load the Rb and Sr into an optical dipole
trap. A quantum degenerate mixture of Rb and Sr will realize
ultracold polar molecules with electron spins and promote research
of new quantum phases.

We acknowledge Prof. T. Kuga for helpful discussions and 
N. Ohtsubo and D. Ikoma for their experimental assistance. 
This research was supported by the Matsuo Foundation and
a Grant-in-Aid for Scientific Research on Innovative Area ``Extreme
quantum world opened by atoms'' (No. 21104005) from the Ministry of
Education, Culture, Sports, Science and Technology, Japan.

\end{document}